\documentclass[conference]{IEEEtran}

\usepackage{cite}
\usepackage{amsmath,amssymb,amsfonts}
\usepackage{graphicx}
\usepackage{booktabs}
\usepackage{algorithm}
\usepackage{algpseudocode}
\usepackage{placeins}
\usepackage[hidelinks]{hyperref}
\graphicspath{{imgs/}}

\begin{document}

\title{Do Speech Tokens Leak Voiceprints? Speaker Inversion Attacks Against End-to-End Speech Language Models}

\author{
\IEEEauthorblockN{Ye Lu, Yihan Yan, Zhaoyang Zhang, Zhitao Ou, Runze Liu, Li Liu, and Shen Wang}
\IEEEauthorblockA{School of Cyberspace Science, Harbin Institute of Technology, Harbin 150001, China\\
Corresponding author: Shen Wang (shen.wang@hit.edu.cn)}
}

\maketitle

\begin{abstract}
End-to-end speech language models increasingly represent user speech with speech tokens rather than relying exclusively on cascaded ASR--LLM--TTS pipelines. Although these tokens support expressive and low-latency spoken interaction, they may also preserve sensitive speaker characteristics. We investigate whether exposed speech tokens leak voiceprints and formulate this risk as a speaker inversion attack. We introduce Audio BERT (AuB), a trainable model that constructs token embeddings from discrete codebooks and aggregates them into speaker-sensitive representations, and propose SpInv, a two-stage inversion method built on AuB to recover embeddings in the space of an attacker-specified speaker encoder. We evaluate Moshi, Higgs3, Kimi-Audio, and Qwen3-Omni using speaker-disjoint protocols on the VoxCeleb dataset. Extensive experiments show that, with only three seconds of frontend output, SpInv achieves cosine similarities above 0.70 in the attacker-specified speaker-encoder space.
\end{abstract}


\section{Introduction}

End-to-end speech language models are emerging as a natural interface for interaction with large models. Unlike cascaded ASR--LLM--TTS systems that convert speech into text for language-model processing, these models directly process speech-modality tokens throughout inference. Such representations preserve prosody, emotion, speaking style, and other acoustic cues, enabling emotion-aware and context-sensitive interaction. They also facilitate bidirectional and full-duplex speech modeling while reducing modality-conversion overhead and response latency.

Directly transmitting user speech to cloud-side models can expose voiceprints and other sensitive acoustic attributes. A split-inference deployment may instead run a tokenizer or speech frontend locally and transmit only its output to the cloud-side model, leaving the downstream provider or an intermediate service with speech tokens rather than the waveform. Speech tokens are more compact and task-oriented than raw audio, but it remains unclear whether they suppress speaker identity or merely encode it in a different form. This motivates our central question: \emph{Do speech tokens in end-to-end speech language models leak voiceprints?}

\begin{figure*}[t]
\centering
\includegraphics[width=\textwidth]{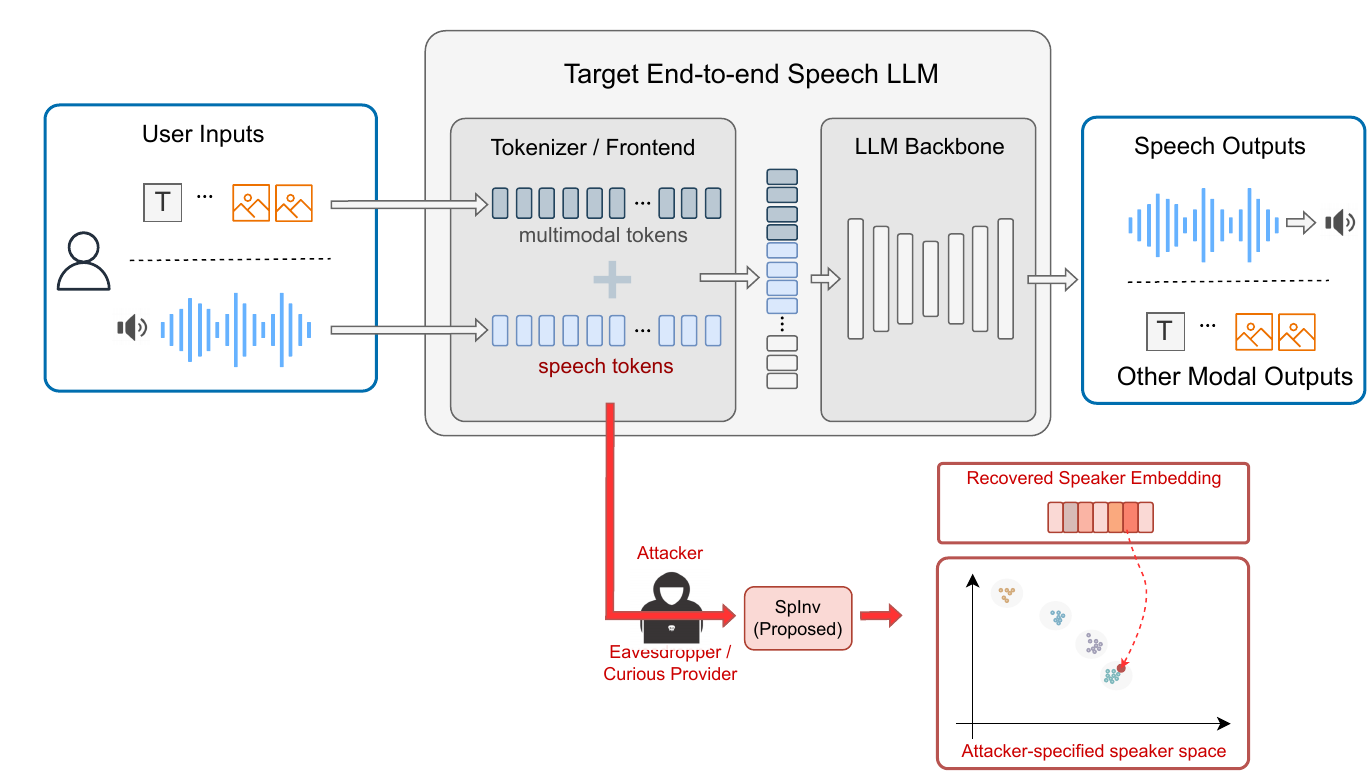}
\caption{Attack scenario. User speech is converted into speech tokens before large-model inference. An eavesdropper or curious provider observing the exposed speech interface uses SpInv to recover an embedding in the space of an attacker-specified speaker encoder.}
\label{fig:teaser}
\end{figure*}

We formulate this privacy question as a Speaker Inversion Attack (SIA) against speech tokens, which extends the concept of a Model Inversion Attack (MIA)~\cite{zhang2020secretrevealer} to speaker privacy in the speech modality. As illustrated in Fig.~\ref{fig:teaser}, the attack arises when an eavesdropper intercepts speech tokens transmitted between the frontend and the downstream model, or when a curious service provider can access them during inference. The adversary then uses the exposed token segment to recover a speaker embedding in the space of a speaker encoder that the adversary specifies. To instantiate this attack, we propose SpInv, a two-stage inversion method that employs Audio BERT (AuB) to aggregate speaker information from speech tokens. AuB maps discrete codebook indices to trainable token embeddings and encodes the resulting sequence with a BERT backbone, while SpInv combines distillation pretraining with discriminative fine-tuning. The trained method recovers a speaker embedding in the attacker-specified speaker-encoder space.

We evaluate four representative speech-model frontends: Moshi, Higgs3, Kimi-Audio, and Qwen3-Omni. The evaluation first tests whether their exposed interfaces retain speaker-discriminative information. It then assesses speaker embedding recovery on speaker-disjoint VoxCeleb datasets followed by analyses of input duration and target speaker space. Our results show that SpInv can recover an embedding in the specified speaker-encoder space from only three seconds of exposed frontend outputs, achieving cosine similarities generally above 0.70 with the target embedding.

The main contributions of this work are as follows:
\begin{itemize}
\item To the best of our knowledge, we present the first systematic study of privacy leakage from speech tokens and extend model inversion in the speech modality to the problem of speaker-identity privacy.
\item We propose SpInv, a two-stage speaker inversion method that employs an AuB model to construct and aggregate token representations and recover embeddings in the space of an attacker-specified speaker encoder.
\item We conduct extensive experiments on four representative target speech frontends, including leakage validation, speaker embedding recovery, and ablations on key factors.
\end{itemize}

\section{Related Work}

\subsection{End-to-End Speech Language Models and Speech Interfaces}

Early spoken interaction systems built around large language models were commonly modular or text-mediated. Representative systems such as SpeechGPT~\cite{zhang2023speechgpt}, Spectron~\cite{nachmani2023spectron}, and PSLM~\cite{mitsui2024pslm} connect speech encoders, language models, and speech-generation modules, retaining structural similarities to cascaded ASR--LLM--TTS pipelines.

Recent systems increasingly retain speech-specific representations throughout a unified inference pipeline. By avoiding repeated conversion between speech and text, speech-token modeling preserves paralinguistic information and supports lower-latency, full-duplex spoken dialogue, which better matches the interactive requirements of speech language models.

Moshi~\cite{defossez2024moshi} is a real-time, full-duplex speech--text model whose Mimi codec maps 24-kHz audio to one semantic and seven residual acoustic token streams at 12.5 frames/s. Higgs3~\cite{bosonai_higgs_audio_tts_v3_2026} is a 4B autoregressive conversational TTS model using eight codebooks at 25 frames/s; its tokenizer is also adopted by the diffusion-based OmniVoice~\cite{zhu2026omnivoice}.
Kimi-Audio~\cite{kimiteam2025kimiaudio} combines a 12.5-Hz semantic-token stream with time-aligned continuous Whisper features. Qwen3-Omni~\cite{xu2025qwen3omni} instead uses an Audio Transformer to emit continuous speech representations at 12.5 Hz. Together, these models cover residual-codebook, hybrid discrete--continuous, and continuous interfaces; AuB handles all three within one voiceprint-leakage attack pipeline.

\subsection{Speaker Representation Learning}

Speaker recognition has evolved from statistical pipelines to deep speaker embeddings. Early methods used GMM-UBM models, joint factor analysis, and i-vectors~\cite{reynolds2000speaker,kenny2007jfa,dehak2011frontend}. Subsequent speaker-verification systems learned discriminative embeddings from large speech corpora~\cite{variani2014dvector,heigold2016endtoend,wan2018ge2e,snyder2018xvectors,chung2018voxceleb2}, with recent work also exploring self-supervised speaker representation learning~\cite{chen_pushing_2023,chen_self-distillation_2025}. ECAPA-TDNN improves TDNN-based verification through channel attention and feature aggregation~\cite{desplanques2020ecapa}; CAM++ uses context-aware masking for efficient speaker modeling~\cite{wang2023campp}; and ERes2Net combines local and global multi-scale features~\cite{chen2023eres2net}. In our threat model, the adversary selects a pretrained speaker encoder to define the target space into which SpInv recovers embeddings.

\subsection{Model Inversion and Audio Privacy Attacks}

Model Inversion Attacks (MIAs) infer sensitive information about model inputs or training data from outputs and intermediate representations. They have been extensively studied in vision, where an adversary may reconstruct representative inputs or identity-related features~\cite{fredrikson2015model,zhang2020secretrevealer,lu_fgmia_2025,li_model_2024}. Related studies in natural language processing have shown that continuous text embeddings can retain enough information to recover substantial properties of their source text~\cite{morris2023text,chen2024multilingual}. These findings demonstrate that replacing raw inputs with learned representations does not necessarily remove private information.

Speech and audio systems expose several distinct privacy surfaces. Prior work has introduced inversion attacks against automatic speaker-recognition systems~\cite{pizzi_introducing_2022} and examined whether speaker identity remains inferable from representations produced by speaker-anonymization mechanisms~\cite{bauer_inference_2025}. Membership-inference studies further investigate whether speech representations reveal the participation of particular users or sensitive attributes in model training~\cite{tsaprazlis_voxguard_2026}. Other attacks exploit released speaker scores to support impersonation~\cite{hwang_scores_2026}, while recent evaluations of audio language models consider whether their outputs disclose sensitive information contained in the audio they process~\cite{wang2026hearsay}. Our work instead targets speech tokens exposed between the speech frontend and the downstream backbone, using a partial token sequence to recover an embedding in the space of an attacker-specified speaker encoder. Unlike waveform-oriented inversion or attacks against speaker-recognition systems, SpInv does not reconstruct the original speech signal or attack the speaker encoder; instead, it recovers an embedding in an attacker-specified speaker space.

\section{Method}

\subsection{Threat Model}

We formalize the attack surface, the adversary's objective, and the knowledge available to the adversary.

\paragraph{Target model}
Let $x$ denote an arbitrary user speech waveform. The target system is an end-to-end speech language model whose speech frontend converts user audio into speech tokens before downstream inference. Let $T$ denote the tokenizer in this frontend, which converts $x$ into a time-indexed speech-token sequence
\begin{equation}
\begin{aligned}
S=T(x)&=[\mathbf{s}_1,\ldots,\mathbf{s}_N],\\
\mathbf{s}_n&=(s_n^{(1)},\ldots,s_n^{(Q)}).
\end{aligned}
\label{eq:frontend-representation}
\end{equation}
Here, $N$ is the sequence length and $Q$ is the number of codebooks. The $q$-th codebook has size $M_q$, and $s_n^{(q)}\in\{1,\ldots,M_q\}$ is its token ID at frame $n$. Some frontends additionally expose time-aligned continuous features $A=[\mathbf{a}_1,\ldots,\mathbf{a}_N]$ alongside the token stream or in place of it. We denote the complete exposed speech representation by $R(x)=(S,A)$, with any unavailable component omitted. We consider deployments in which these frontend outputs are accessible to an intermediate service or transmitted to the downstream model.

\paragraph{Adversary goal}
The adversary is an eavesdropper observing the exposed speech interface or a curious service provider with access to it during inference, consistent with speech-privacy threat models that assume access to intermediate representations~\cite{tsaprazlis_voxguard_2026,wang2026hearsay}. Let $F$ be a frozen speaker encoder selected by the attacker that maps a waveform to a target speaker embedding
\begin{equation}
\mathbf{z}=F(x)\in\mathbb{R}^{d_z},
\label{eq:target-speaker-embedding}
\end{equation}
where $d_z$ is the embedding dimension. By selecting $F$, the attacker specifies the target speaker-embedding space to be recovered. Rather than reconstructing the original waveform or recovering the target model's parameters, the adversary observes a randomly cropped contiguous segment $\widetilde{R}$ of $R(x)$ and seeks to recover $\mathbf{z}$ without access to the victim's waveform at attack time. A recovery model $G_{\theta}$ with parameters $\theta$ produces $\hat{\mathbf{z}}=G_{\theta}(\widetilde{R})\in\mathbb{R}^{d_z}$. We formulate the Speaker Inversion Attack as
\begin{equation}
\theta^{*}=\arg\min_{\theta}\;
\mathcal{L}\bigl(G_{\theta}(\widetilde{R}),\mathbf{z}\bigr),
\label{eq:attack-objective}
\end{equation}
where $\mathcal{L}$ measures the discrepancy between the recovered and target embeddings in the speaker space defined by $F$, and $\theta^{*}$ denotes the optimized recovery-model parameters.

\paragraph{Adversary knowledge and capabilities}
The adversary is assumed to have three capabilities. First, it has query access to the target speech frontend $T$: it can submit an utterance and observe the corresponding frontend output $R(x)$, but it does not access the internal parameters of the target model. Second, for a token-based interface, it can observe the number of codebooks $Q$ and their vocabulary sizes $\{M_q\}_{q=1}^{Q}$; the learned codebook embedding values are neither known nor required. Third, it can obtain a public auxiliary speech dataset $\mathcal{D}$ and use the queried frontend outputs to train the attack model. The adversary selects and has access to the frozen speaker encoder $F$; this choice determines the desired speaker-embedding space into which $G_{\theta}$ must recover embeddings. In our experiments, $F$ is instantiated by a publicly available pretrained speaker encoder.

\subsection{Overview of SpInv}

\begin{figure*}[t]
\centering
\includegraphics[width=\textwidth]{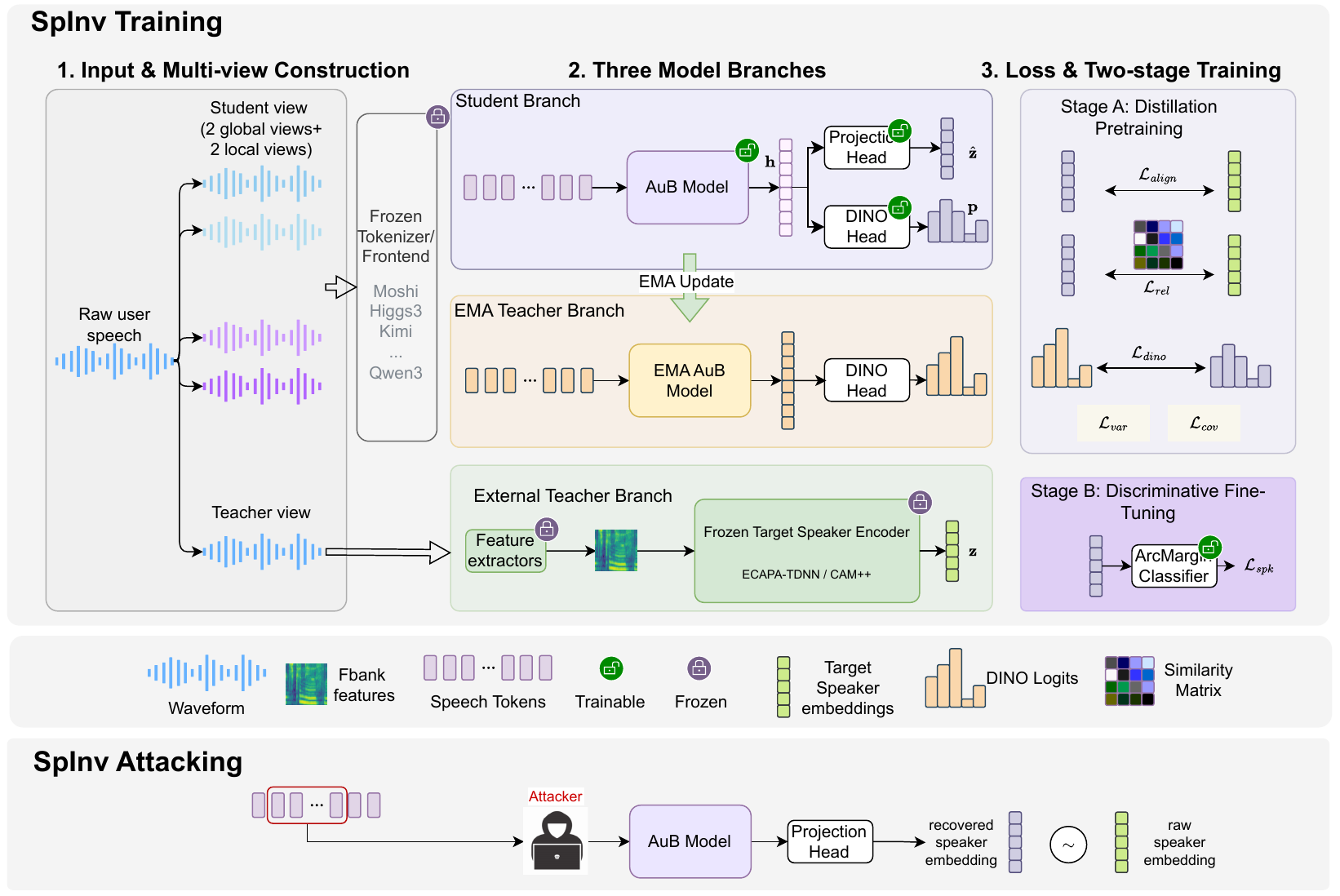}
\caption{Overview of SpInv. The training phase learns a recovery model from frontend outputs via two-stage optimization; the attack phase recovers a speaker embedding from an intercepted token slice.}
\label{fig:overview}
\end{figure*}

As illustrated in Fig.~\ref{fig:overview}, SpInv encompasses both a training phase and an attack phase. The training phase learns a recovery model $G_{\theta}$ from the outputs of the target speech frontend through a two-stage strategy that combines distillation pretraining and discriminative fine-tuning. In the attack phase, the adversary intercepts only a slice of exposed speech tokens from the victim's frontend stream; the trained AuB model and projection head then recover a speaker embedding $\hat{\mathbf{z}}$ in the selected speaker space directly from this token slice, without requiring the original waveform.

\subsection{AuB Model Structure}

AuB is the trainable speaker information extractor used by SpInv. It comprises a codebook embedding layer, a feature fusion layer, and a BERT backbone. For codebook $q$ with vocabulary size $M_q$, AuB maintains a trainable lookup table $\mathbf{E}^{(q)}\in\mathbb{R}^{M_q\times d}$. The token IDs at frame $n$ are mapped to the shared embedding space and added across codebooks:
\begin{equation}
\mathbf{e}_n=\sum_{q=1}^{Q}\mathbf{E}^{(q)}[s_n^{(q)}],
\qquad \mathbf{e}_n\in\mathbb{R}^{d}.
\label{eq:codebook-embedding}
\end{equation}
The lookup tables belong to AuB and are learned independently of the unknown codebook embeddings used by the target model.

Since certain target frontends additionally expose the time-aligned continuous feature $\mathbf{a}_n$ at each frame as supplementary acoustic information, AuB incorporates a feature fusion layer for compatibility. An MLP projects this feature to the shared embedding space as $\mathbf{u}_n=\operatorname{MLP}(\mathbf{a}_n)\in\mathbb{R}^{d}$. The fused representation at frame $n$ is obtained by element-wise addition:
\begin{equation}
\mathbf{r}_n=\mathbf{e}_n+\mathbf{u}_n.
\label{eq:feature-fusion}
\end{equation}
This design accommodates the native exposed outputs of all target frontends within a single architecture.

Similar to ViT~\cite{dosovitskiy2021image}, AuB prepends a learnable CLS token to $[\mathbf{r}_1,\ldots,\mathbf{r}_L]$ and processes the sequence with a BERT backbone~\cite{devlin2019bert}, where $L$ is the length of the input segment. The CLS output $\mathbf{h}$ is converted into a fixed 512-dimensional AuB output $\mathbf{o}\in\mathbb{R}^{d_a}$, where $d_a=512$. The projection head is a two-layer MLP with LayerNorm~\cite{ba2016layer} and GELU~\cite{hendrycks2016gelu} that maps $\mathbf{o}$ to the recovered embedding $\hat{\mathbf{z}}\in\mathbb{R}^{d_z}$. Accordingly, the recovery model $G_{\theta}$ in~\eqref{eq:attack-objective} comprises AuB and the projection head, with $\theta=(\theta_A,\theta_P)$ denoting their respective parameters. During training, $G_{\theta}$ also exposes $\mathbf{h}$ and $\mathbf{o}$ for auxiliary objectives. A separate DINO head $D_{\phi_D}$ uses a two-layer MLP projector with GELU, followed by bottleneck LayerNorm, $\ell_2$ normalization, and a bias-free prototype layer to produce logits $\boldsymbol{\ell}=D_{\phi_D}(\mathbf{o})$. The DINO head is optimized only as an auxiliary training component and is discarded after training; it is therefore not part of $G_{\theta}$ or the final recovery parameters $\theta^{*}$. AuB, the projection head, and the auxiliary DINO head are randomly initialized and trained jointly.

\subsection{SpInv Training and Attack}

\paragraph{Training phase}
As shown in the training phase of Fig.~\ref{fig:overview}, SpInv organizes three parallel branches. All five views are constructed at the waveform level for each utterance: two local views $\mathcal{V}_s=\{l_1,l_2\}$, two global views $\mathcal{V}_g=\{g_1,g_2\}$, and one teacher view $x_i^t$. The student model receives only the two local views through the frozen target frontend $R$ and $G_{\theta}$, producing recovered embeddings $\hat{\mathbf{z}}$, CLS representations $\mathbf{h}$, AuB outputs $\mathbf{o}$, and student logits from $D_{\phi_D}$. The EMA teacher model, comprising EMA copies of AuB and the DINO head, receives only the two global views and generates self-distillation targets. The frozen teacher model $F$ receives only the teacher view and produces the target embedding $\mathbf{z}_i=F(x_i^t)$. The loss block in Fig.~\ref{fig:overview} shows that $L_{\mathrm{align}}$ and $L_{\mathrm{rel}}$ operate on embeddings recovered from the local views, $L_{\mathrm{dino}}$ matches student distributions from the local views against EMA teacher targets from the global views, and $L_{\mathrm{var}}$ and $L_{\mathrm{cov}}$ regularize representations from the local views. In Stage~B, an ArcMargin classifier adds speaker-discriminative supervision on the CLS representations from the local views.

\textit{Stage A: Distillation pretraining.} Stage~A combines alignment to the target speaker space, relational matching of batch geometry, and EMA-based self-distillation across views. Variance and covariance regularization following VICReg~\cite{bardes2022vicreg} prevent representational collapse. Its objective is
\begin{equation}
L_A=\lambda_1L_{\mathrm{align}}+\lambda_2L_{\mathrm{rel}}+\lambda_3L_{\mathrm{dino}}
+\lambda_4L_{\mathrm{var}}+\lambda_5L_{\mathrm{cov}},
\label{eq:stage-a-objective}
\end{equation}
where $\lambda_j\geq 0$ are loss weights.

For a mini-batch of $B$ utterances, let $V_s=|\mathcal{V}_s|=2$, let $\hat{\mathbf{z}}_{i,v}$ be the recovered embedding of utterance $i$ under local view $v\in\mathcal{V}_s$, and let $\mathbf{z}_i$ be the target embedding obtained from its teacher view. The alignment loss is
\begin{equation}
L_{\mathrm{align}}=\frac{1}{BV_s}\sum_{i=1}^{B}\sum_{v\in\mathcal{V}_s}
\bigl(1-\cos(\hat{\mathbf{z}}_{i,v},\mathbf{z}_i)\bigr).
\label{eq:alignment-loss}
\end{equation}
The relational loss aligns pairwise speaker geometry within the batch:
\begin{equation}
L_{\mathrm{rel}}=\frac{1}{V_sB(B-1)}\sum_{v\in\mathcal{V}_s}\sum_{i\ne j}
\bigl(\cos(\hat{\mathbf{z}}_{i,v},\hat{\mathbf{z}}_{j,v})-
\cos(\mathbf{z}_i,\mathbf{z}_j)\bigr)^2.
\label{eq:relational-loss}
\end{equation}

For self-distillation, the valid teacher--student pairs form the Cartesian product $\mathcal{P}=\mathcal{V}_g\times\mathcal{V}_s$. The EMA teacher produces centered and sharpened prototype distributions $q_{i,u}$ from global view $u\in\mathcal{V}_g$, while the student produces prototype distributions $p_{i,v}$ from local view $v\in\mathcal{V}_s$. The DINO loss~\cite{caron2021dino} is
\begin{equation}
L_{\mathrm{dino}}=-\frac{1}{B|\mathcal{P}|}
\sum_{(u,v)\in\mathcal{P}}\sum_{i=1}^{B}\sum_{k=1}^{K}
\operatorname{sg}(q_{i,u,k})\log p_{i,v,k},
\label{eq:dino-loss}
\end{equation}
where $K$ is the number of prototypes and $\operatorname{sg}(\cdot)$ denotes stop-gradient.

Let $H^{(v)}\in\mathbb{R}^{B\times d_h}$ contain the student representations used for regularization under view $v$. With $\gamma>0$ denoting the target standard deviation and $\epsilon>0$ ensuring numerical stability, the view-averaged anti-collapse terms are
\begin{equation}
L_{\mathrm{var}}=\frac{1}{V_sd_h}\sum_{v\in\mathcal{V}_s}\sum_{j=1}^{d_h}
\max\bigl(0,\gamma-\sqrt{\operatorname{Var}(H^{(v)}_{:,j})+\epsilon}\bigr),
\label{eq:variance-loss}
\end{equation}
\begin{equation}
L_{\mathrm{cov}}=\frac{1}{V_sd_h}\sum_{v\in\mathcal{V}_s}\sum_{j\ne j'}
\bigl(\operatorname{Cov}(H^{(v)})_{j,j'}\bigr)^2.
\label{eq:covariance-loss}
\end{equation}

\textit{Stage B: Discriminative fine-tuning.} Stage~B initializes $G_{\theta}$ and the auxiliary DINO head from the selected Stage~A checkpoint and adds speaker-discriminative supervision. An ArcMargin classifier~\cite{deng2019arcface} is applied to $\mathbf{h}$, while lower-weight versions of the Stage~A losses preserve the learned target-space structure:
\begin{equation}
L_B=\mu_1L_{\mathrm{spk}}+\mu_2L_{\mathrm{align}}+\mu_3L_{\mathrm{rel}}
+\mu_4L_{\mathrm{dino}}+\mu_5L_{\mathrm{var}}+\mu_6L_{\mathrm{cov}},
\label{eq:stage-b-objective}
\end{equation}
where $\mu_j\geq 0$ are loss weights. Let $C_{\mathrm{spk}}$ be the number of speakers in the auxiliary training set, with $c\in\{1,\ldots,C_{\mathrm{spk}}\}$. For normalized $\bar{\mathbf{h}}_{i,v}$ and classifier weight $\bar{\mathbf{w}}_c$, define $\alpha_{i,v,c}=\arccos(\bar{\mathbf{h}}_{i,v}^{\top}\bar{\mathbf{w}}_c)$ and the ArcMargin logit $\beta_{i,v,c}=s\cos(\alpha_{i,v,c}+m\mathbb{I}[c=y_i])$. The speaker-classification loss over the local views is
\begin{equation}
L_{\mathrm{spk}}=-\frac{1}{BV_s}\sum_{i=1}^{B}\sum_{v\in\mathcal{V}_s}
\log\frac{e^{\beta_{i,v,y_i}}}
{\sum_{c=1}^{C_{\mathrm{spk}}}e^{\beta_{i,v,c}}},
\label{eq:speaker-loss}
\end{equation}
where $y_i$ is the speaker label, $m$ is the angular margin, and $s$ is the logit scale. Stage~B retains the same assignment as Stage~A: two local views to the student model, two global views to the EMA teacher model, and one teacher view to the teacher model.

\paragraph{Attack phase}
As illustrated in the attack phase of Fig.~\ref{fig:overview}, the adversary does not access the victim's waveform but only intercepts a contiguous slice of exposed speech tokens (and any accompanying interface features) from the victim's frontend output stream. This token slice is fed directly to the trained recovery model $G_{\theta}$, comprising AuB and the projection head, to yield $\hat{\mathbf{z}}$ in the speaker space specified by $F$. The auxiliary DINO head and ArcMargin classifier are not used during the attack. As demonstrated experimentally, the recovered speaker embedding $\hat{\mathbf{z}}$ has high cosine similarity to the raw speaker embedding $\mathbf{z}=F(x)$ extracted by $F$ when the waveform is available for evaluation.

\section{Experiments}

\subsection{Experimental Setup}

\paragraph{Target models}
We evaluate SpInv against four representative speech-model frontends with their native exposed outputs. Moshi~\cite{defossez2024moshi} employs the Mimi neural audio codec, which exposes one semantic and seven residual acoustic token streams at 12.5 frames/s. Higgs3~\cite{bosonai_higgs_audio_tts_v3_2026} uses an eight-codebook tokenizer operating at 25 frames/s; this tokenizer is also used as the audio frontend in OmniVoice~\cite{zhu2026omnivoice}. Kimi-Audio~\cite{kimiteam2025kimiaudio} outputs a semantic-token stream together with time-aligned continuous Whisper features, which provide complementary acoustic information. Qwen3-Omni~\cite{xu2025qwen3omni} uses a continuous Audio Transformer frontend that emits frame-level feature vectors at 12.5 Hz. The first three models expose discrete speech-token interfaces and constitute the core focus of this study; Qwen3-Omni is additionally evaluated to demonstrate that SpInv also applies to purely continuous frontend features, broadening the evaluation coverage. SpInv uses the exposed output of each frontend directly. For each frontend, input waveforms are resampled to its native sampling rate; all frontend parameters remain frozen throughout attack training.

\paragraph{Datasets}
We use the VoxCeleb corpora, which are widely adopted benchmarks for speaker recognition research. VoxCeleb1~\cite{nagrani2017voxceleb} contains over 150,000 utterances from 1,251 speakers, totaling approximately 352 hours of YouTube speech. Under its standard verification partition, the development set contains 1,211 speakers and the test set contains 40 speakers, with no speaker identity overlap. VoxCeleb2~\cite{chung2018voxceleb2} is substantially larger, comprising over 1.1 million utterances, totaling roughly 2,442 hours, from 6,112 speakers. Its standard partition contains 5,994 development speakers and 118 test speakers, which are also identity-disjoint. Both corpora cover diverse acoustic conditions, speaking styles, and demographic backgrounds.

We follow the standard VoxCeleb1 and VoxCeleb2 evaluation protocols. Privacy leakage validation uses the official VoxCeleb1 verification split~\cite{nagrani2017voxceleb}. For speaker embedding recovery, SpInv is trained on VoxCeleb2-dev and evaluated on the speaker-disjoint VoxCeleb2-test and VoxCeleb1 sets. The speakers in VoxCeleb1 and VoxCeleb2-dev do not overlap. Unless stated otherwise, the subsequent ablations use the Moshi tokenizer with VoxCeleb2-dev $\rightarrow$ VoxCeleb2-test.

\paragraph{Evaluation method}
The frozen teacher model extracts target speaker embeddings from the teacher view, while the student model receives only target-frontend outputs derived from the local views. Let $\mathbf{z}_i$ be the target embedding extracted by the frozen speaker encoder $F$ and $\hat{\mathbf{z}}_i$ be the embedding recovered by SpInv. CosSim measures their average cosine similarity:
\begin{equation}
\mathrm{CosSim}=\frac{1}{N}\sum_{i=1}^{N}
\frac{\hat{\mathbf{z}}_i^{\top}\mathbf{z}_i}
{\|\hat{\mathbf{z}}_i\|_2\|\mathbf{z}_i\|_2}.
\label{eq:cosine-similarity}
\end{equation}
We also report KL after converting the two embeddings into distributions over dimensions, $P_i=\operatorname{softmax}(\mathbf{z}_i)$ and $\hat{P}_i=\operatorname{softmax}(\hat{\mathbf{z}}_i)$:
\begin{equation}
\mathrm{KL}=\frac{1}{N}\sum_{i=1}^{N}\sum_k
P_{i,k}\log\frac{P_{i,k}+\epsilon}{\hat{P}_{i,k}+\epsilon},
\label{eq:kl-divergence}
\end{equation}
where $\epsilon$ is a small constant. Accuracy (Acc.), equal error rate (EER), and normalized minDCF measure speaker-verification performance. For minDCF, we use $P_{\mathrm{target}}=0.01$, $C_{\mathrm{miss}}=1$, and $C_{\mathrm{fa}}=1$. The reported threshold is the corresponding optimal cosine-score decision boundary used to compute Acc.

\paragraph{Target speaker space}
The target speaker space is defined by the frozen encoder $F$ selected by the attacker, as formalized in~\eqref{eq:target-speaker-embedding}. In the main experiments, we adopt ECAPA-TDNN~\cite{desplanques2020ecapa} with embedding dimensionality $d_z=192$ as the default target space. To assess the robustness of SpInv across different speaker representations, we also evaluate three alternative target spaces: ERes2Net~\cite{chen2023eres2net} ($d_z=192$), which captures multi-scale speaker features, and CAM++~\cite{wang2023campp} at two embedding dimensionalities ($d_z=192$ and $d_z=512$), which employs context-aware masking for efficient speaker modeling. All target encoders are publicly available, pretrained, and kept frozen during attack training.

\paragraph{Default hyperparameter configuration}
The AuB codebook embedding layer uses the same number of codebooks and vocabulary sizes as observed in the target frontend. The feature fusion layer uses a one-layer MLP. The projection and DINO heads follow the structures described in Section~III-C. AuB uses a BERT hidden dimension of $d_h=384$ and produces a fixed-dimensional output with $d_a=512$. The projection head maps this 512-dimensional output to $d_z$, the dimensionality of the selected target speaker encoder, such as $d_z=192$ for the default ECAPA-TDNN. The complete AuB model is trained from random initialization using Adam with a batch size of 96 (which occupies approximately 20\,GB of GPU memory), while the target frontend and speaker encoder remain frozen. Stage~A runs for 60 epochs with a learning rate of $4\times10^{-4}$ and weights $\lambda_1=1.0$, $\lambda_2=0.5$, $\lambda_3=1.0$, $\lambda_4=25.0$, and $\lambda_5=1.0$ for $L_{\mathrm{align}}$, $L_{\mathrm{rel}}$, $L_{\mathrm{dino}}$, $L_{\mathrm{var}}$, and $L_{\mathrm{cov}}$, respectively. Stage~B runs for 40 epochs with a learning rate of $1\times10^{-4}$ and weights $\mu_1=0.2$, $\mu_2=0.8$, $\mu_3=0.4$, $\mu_4=0.4$, $\mu_5=10.0$, and $\mu_6=0.5$ for $L_{\mathrm{spk}}$, $L_{\mathrm{align}}$, $L_{\mathrm{rel}}$, $L_{\mathrm{dino}}$, $L_{\mathrm{var}}$, and $L_{\mathrm{cov}}$. The ArcMargin margin is linearly increased from 0 to 0.2 between epochs 10 and 20 of Stage~B. The DINO center momentum is 0.9, and the EMA momentum follows a cosine schedule from 0.996 to 0.9995.

\subsection{Privacy Leakage Validation}

Before evaluating inversion, we determine whether the exposed speech interfaces retain speaker-discriminative information. We train AuB with speaker labels on the VoxCeleb1 development partition and evaluate its representations on the speaker-disjoint VoxCeleb1 verification test partition. Random three-second cropping is the only data augmentation. Table~\ref{tab:leakage} reports this diagnostic experiment for all four target models.

\begin{table}[t]
\caption{Speaker-information leakage on the speaker-disjoint VoxCeleb1 verification split.}
\label{tab:leakage}
\centering
\begin{tabular}{lcc}
\toprule
Target model & EER $\downarrow$ & minDCF $\downarrow$ \\
\midrule
Moshi & 0.0160 & 0.0512 \\
Higgs3 & 0.0066 & 0.0132 \\
Kimi-Audio & 0.0034 & 0.0171 \\
Qwen3-Omni & 0.0236 & 0.0836 \\
\bottomrule
\end{tabular}
\end{table}

All four exposed interfaces retain strong speaker-discriminative information under the speaker-disjoint protocol. Kimi-Audio achieves the lowest EER of 0.0034, followed by Higgs3 at 0.0066 and Moshi at 0.0160, while Qwen3-Omni also exhibits substantial leakage with an EER of 0.0236. These results confirm that the frontend representations preserve speaker information that generalizes to unseen identities.

\subsection{Speaker Embedding Recovery}

We next evaluate whether SpInv can recover embeddings in the frozen ECAPA-TDNN speaker space for previously unseen speakers. By default, the attack observes three seconds of exposed frontend output. Table~\ref{tab:recovery-v2} reports results on VoxCeleb2-test, and Table~\ref{tab:recovery-v1} reports results on VoxCeleb1; both use VoxCeleb2-dev for attack training.

\begin{table}[t]
\caption{Speaker embedding recovery from VoxCeleb2-dev to VoxCeleb2-test.}
\label{tab:recovery-v2}
\centering
\resizebox{\columnwidth}{!}{
\begin{tabular}{lcccccc}
\toprule
Target model & CosSim $\uparrow$ & KL $\downarrow$ & Acc $\uparrow$ & EER $\downarrow$ & minDCF $\downarrow$ & Thres. \\
\midrule
Moshi & 0.7136 & 0.3346 & 0.9701 & 0.0305 & 0.0885 & 0.3287 \\
Higgs3 & 0.7527 & 0.2858 & 0.9828 & 0.0174 & 0.0627 & 0.3870 \\
Kimi-Audio & 0.7536 & 0.2919 & 0.9825 & 0.0178 & 0.0723 & 0.4215 \\
Qwen3-Omni & 0.6683 & 0.3257 & 0.9624 & 0.0378 & 0.1124 & 0.3269 \\
\bottomrule
\end{tabular}
}
\end{table}

\begin{table}[t]
\caption{Speaker embedding recovery from VoxCeleb2-dev to VoxCeleb1.}
\label{tab:recovery-v1}
\centering
\resizebox{\columnwidth}{!}{
\begin{tabular}{lcccccc}
\toprule
Target model & CosSim $\uparrow$ & KL $\downarrow$ & Acc $\uparrow$ & EER $\downarrow$ & minDCF $\downarrow$ & Thres. \\
\midrule
Moshi & 0.7037 & 0.4257 & 0.9395 & 0.0606 & 0.1245 & 0.2833  \\
Higgs3 & 0.7060 & 0.3540 & 0.9513 & 0.0489 & 0.1016 & 0.3208  \\
Kimi-Audio & 0.7359 & 0.3395 & 0.9449 & 0.0554 & 0.0991 & 0.3422  \\
Qwen3-Omni & 0.6594 & 0.3991 & 0.9219 & 0.0783 & 0.1335 & 0.2240  \\
\bottomrule
\end{tabular}
}
\end{table}

Across both speaker-disjoint evaluation sets, Higgs3 and Kimi-Audio provide the strongest overall recovery, followed by Moshi, while Qwen3-Omni is the most difficult frontend to attack. All four frontends perform worse on VoxCeleb1 than on VoxCeleb2-test. Because SpInv is trained on VoxCeleb2-dev, this consistent decline may result from a slight distribution mismatch between the two evaluation datasets, while the relative performance trend remains stable.

\subsection{Ablation Study}

We examine the effects of attack-time input duration and the target speaker space using Moshi with VoxCeleb2-dev $\rightarrow$ VoxCeleb2-test.

\paragraph{Input duration}
We vary the observed segment duration from 1 to 10 seconds while using ECAPA-TDNN as the target speaker encoder. Table~\ref{tab:duration} reports how the amount of exposed speech affects embedding recovery.

\begin{table}[t]
\caption{Effect of attack-time input duration using Moshi (VoxCeleb2-dev $\rightarrow$ VoxCeleb2-test).}
\label{tab:duration}
\centering
\resizebox{\columnwidth}{!}{
\begin{tabular}{lcccccc}
\toprule
Duration & CosSim $\uparrow$ & KL $\downarrow$ & Acc $\uparrow$ & EER $\downarrow$ & minDCF $\downarrow$ & Thres. \\
\midrule
1\,s & 0.4887 & 0.4353 & 0.8656 & 0.1346 & 0.6670 & 0.1868 \\
2\,s & 0.6677 & 0.3537 & 0.9559 & 0.0444 & 0.4050 & 0.2697 \\
3\,s & 0.7136 & 0.3346 & 0.9701 & 0.0305 & 0.0885 & 0.3287 \\
5\,s & 0.7642 & 0.3176 & 0.9849 & 0.0153 & 0.0878 & 0.3694 \\
8\,s & 0.7949 & 0.3131 & 0.9892 & 0.0110 & 0.0707 & 0.3723 \\
10\,s & 0.8065 & 0.3116 & 0.9878 & 0.0124 & 0.0757 & 0.4132 \\
\bottomrule
\end{tabular}
}
\end{table}

Shorter observations reduce speaker-embedding recovery performance, with the largest degradation occurring at one second. Starting from three seconds, the attack achieves substantially stronger recovery. Longer observations generally improve CosSim, although the verification gains become smaller and mildly non-monotonic beyond eight seconds, indicating that the attack approaches saturation.

\paragraph{Target speaker space}
We evaluate recovery into several attacker-specified speaker-encoder spaces instantiated with publicly available models, including ECAPA-TDNN, ERes2Net~\cite{chen2023eres2net}, and CAM++~\cite{wang2023campp}. Table~\ref{tab:targetspace} reports the results for each target space.

\begin{table}[t]
\caption{Effect of the target speaker space using Moshi (VoxCeleb2-dev $\rightarrow$ VoxCeleb2-test).}
\label{tab:targetspace}
\centering
\resizebox{\columnwidth}{!}{
\begin{tabular}{lcccccc}
\toprule
Speaker encoder & Dim & CosSim $\uparrow$ & Acc $\uparrow$ & EER $\downarrow$ & minDCF $\downarrow$ & Thres. \\
\midrule
ECAPA-TDNN & 192 & 0.7136 & 0.9701 & 0.0305 & 0.0885 & 0.3287 \\
ERes2Net & 192 & 0.7562 & 0.9559 & 0.0443 & 0.0882 & 0.3389 \\
CAM++ & 192 & 0.7658 & 0.9811 & 0.0191 & 0.0957 & 0.3074 \\
CAM++ & 512 & 0.7167 & 0.9867 & 0.0135 & 0.1044 & 0.3040 \\
\bottomrule
\end{tabular}
}
\end{table}

SpInv achieves effective recovery across all four target speaker spaces. CAM++ with 192-dimensional embeddings obtains the highest CosSim of 0.7658, whereas its 512-dimensional variant achieves the lowest EER of 0.0135 and the highest Acc. of 0.9867. These results indicate that the attack is not restricted to a specific speaker encoder or embedding dimensionality.

\section{Conclusion}

This paper investigates whether speech tokens exposed by end-to-end speech language models leak sufficient voiceprint information to recover speaker embeddings. We formulate this risk as a speaker inversion attack and propose SpInv, a two-stage method that uses AuB to construct and aggregate discrete or continuous frontend representations and recover embeddings in an attacker-specified speaker space. We evaluate SpInv against Moshi, Higgs3, Kimi-Audio, and Qwen3-Omni using speaker-disjoint protocols on VoxCeleb2-test and VoxCeleb1. The results show that exposed frontend representations retain substantial speaker information: with only three seconds of observation, SpInv achieves strong recovery on unseen speakers, and the three token-based frontends generally reach cosine similarities above 0.70.

This study has two main limitations. First, it focuses on recovering speaker embeddings rather than reconstructing waveforms that reproduce a victim's acoustic characteristics. Second, the experiments are limited to the VoxCeleb corpora rather than a broader range of unconstrained in-the-wild speech datasets. Future work will investigate privacy-preserving tokenizers that better suppress voiceprint information while retaining the content needed by downstream speech-language tasks. We also plan to study stronger attacks that recover waveforms from exposed speech representations.

\section{Ethical Statement}

This work is intended to identify and quantify voiceprint-privacy risks in exposed speech representations so that safer speech interfaces can be developed. All experiments use publicly available research datasets and pretrained models, and results are reported only in aggregate. We do not attempt to identify, contact, or impersonate individuals represented in the evaluation data, nor do we reconstruct their speech waveforms. Nevertheless, speaker inversion could be misused for unauthorized profiling or identity inference. Any release or application of such techniques should follow applicable dataset licenses, consent requirements, and data-protection regulations, and should incorporate access controls that prevent attacks on non-consenting users. Our findings are presented to motivate privacy-preserving tokenizers and stronger safeguards for deployed speech-language systems.

\bibliographystyle{IEEEtran}
\bibliography{refs}

@inproceedings{bauer_inference_2025,
  title     = {Inference attacks for x-vector speaker anonymization},
  author    = {Bauer, Luke A. and Bao, Wenxuan and Jadhav, Malvika and Bindschaedler, Vincent},
  booktitle = {Proc. IEEE Secur. Priv. Workshops (SPW)},
  pages     = {152--159},
  year      = {2025}
}

@inproceedings{bardes2022vicreg,
  title     = {{VICReg}: Variance-invariance-covariance regularization for self-supervised learning},
  author    = {Bardes, Adrien and Ponce, Jean and LeCun, Yann},
  booktitle = {Proc. Int. Conf. Learn. Represent. (ICLR)},
  year      = {2022}
}

@inproceedings{caron2021dino,
  title     = {Emerging properties in self-supervised vision transformers},
  author    = {Caron, Mathilde and Touvron, Hugo and Misra, Ishan and J{\'e}gou, Herv{\'e} and Mairal, Julien and Bojanowski, Piotr and Joulin, Armand},
  booktitle = {Proc. IEEE/CVF Int. Conf. Comput. Vis. (ICCV)},
  pages     = {9650--9660},
  year      = {2021}
}

@misc{bosonai_higgs_audio_tts_v3_2026,
  title        = {{Higgs TTS 3}: Conversational speech for voice {AI} from {Boson AI}},
  author       = {{Boson AI}},
  howpublished = {\url{https://huggingface.co/bosonai/higgs-tts-3-4b}},
  year         = {2026}
}

@inproceedings{chen_pushing_2023,
  title     = {Pushing the limits of self-supervised speaker verification using regularized distillation framework},
  author    = {Chen, Yafeng and Zheng, Siqi and Wang, Hui and Cheng, Luyao and Chen, Qian},
  booktitle = {Proc. IEEE Int. Conf. Acoust. Speech Signal Process. (ICASSP)},
  pages     = {1--5},
  year      = {2023}
}

@inproceedings{chen_self-distillation_2025,
  title     = {Self-distillation prototypes network: Learning robust speaker representations without supervision},
  author    = {Chen, Yafeng and Zheng, Siqi and Wang, Hui and Cheng, Luyao and Chen, Qian and Deng, Chong and Zhang, Shiliang and Wang, Wen},
  booktitle = {Proc. IEEE Int. Conf. Acoust. Speech Signal Process. (ICASSP)},
  pages     = {1--5},
  year      = {2025}
}

@inproceedings{chen2023eres2net,
  title     = {An enhanced {Res2Net} with local and global feature fusion for speaker verification},
  author    = {Chen, Yafeng and Zheng, Siqi and Wang, Hui and Cheng, Luyao and Chen, Qian and Qi, Jiajun},
  booktitle = {Proc. Annu. Conf. Int. Speech Commun. Assoc. (Interspeech)},
  pages     = {2228--2232},
  year      = {2023}
}

@inproceedings{chen2024multilingual,
  title     = {Text embedding inversion security for multilingual language models},
  author    = {Chen, Yiyi and Lent, Heather and Bjerva, Johannes},
  booktitle = {Proc. Annu. Meet. Assoc. Comput. Linguist. (ACL)},
  pages     = {7808--7827},
  year      = {2024}
}

@inproceedings{chung2018voxceleb2,
  title     = {{VoxCeleb2}: Deep speaker recognition},
  author    = {Chung, Joon Son and Nagrani, Arsha and Zisserman, Andrew},
  booktitle = {Proc. Annu. Conf. Int. Speech Commun. Assoc. (Interspeech)},
  pages     = {1086--1090},
  year      = {2018}
}

@article{dehak2011frontend,
  title   = {Front-end factor analysis for speaker verification},
  author  = {Dehak, Najim and Kenny, Patrick J. and Dehak, Reda and Dumouchel, Pierre and Ouellet, Pierre},
  journal = {IEEE Trans. Audio Speech Lang. Process.},
  volume  = {19},
  number  = {4},
  pages   = {788--798},
  year    = {2011}
}

@article{defossez2024moshi,
  title   = {Moshi: A speech-text foundation model for real-time dialogue},
  author  = {D{\'e}fossez, Alexandre and Mazar{\'e}, Laurent and Orsini, Manu and Royer, Am{\'e}lie and P{\'e}rez, Patrick and J{\'e}gou, Herv{\'e} and Grave, Edouard and Zeghidour, Neil},
  journal = {arXiv preprint arXiv:2410.00037},
  year    = {2024}
}

@inproceedings{deng2019arcface,
  title     = {{ArcFace}: Additive angular margin loss for deep face recognition},
  author    = {Deng, Jiankang and Guo, Jia and Xue, Niannan and Zafeiriou, Stefanos},
  booktitle = {Proc. IEEE/CVF Conf. Comput. Vis. Pattern Recognit. (CVPR)},
  pages     = {4690--4699},
  year      = {2019}
}

@inproceedings{desplanques2020ecapa,
  title     = {{ECAPA-TDNN}: Emphasized channel attention, propagation and aggregation in {TDNN} based speaker verification},
  author    = {Desplanques, Brecht and Thienpondt, Jenthe and Demuynck, Kris},
  booktitle = {Proc. Annu. Conf. Int. Speech Commun. Assoc. (Interspeech)},
  pages     = {3830--3834},
  year      = {2020}
}

@inproceedings{devlin2019bert,
  title     = {{BERT}: Pre-training of deep bidirectional transformers for language understanding},
  author    = {Devlin, Jacob and Chang, Ming-Wei and Lee, Kenton and Toutanova, Kristina},
  booktitle = {Proc. North Am. Chapter Assoc. Comput. Linguist.: Hum. Lang. Technol. (NAACL-HLT)},
  pages     = {4171--4186},
  year      = {2019}
}

@inproceedings{dosovitskiy2021image,
  title     = {An image is worth 16x16 words: Transformers for image recognition at scale},
  author    = {Dosovitskiy, Alexey and Beyer, Lucas and Kolesnikov, Alexander and Weissenborn, Dirk and Zhai, Xiaohua and Unterthiner, Thomas and Dehghani, Mostafa and Minderer, Matthias and Heigold, Georg and Gelly, Sylvain and Uszkoreit, Jakob and Houlsby, Neil},
  booktitle = {Proc. Int. Conf. Learn. Represent. (ICLR)},
  year      = {2021}
}

@article{hendrycks2016gelu,
  title   = {Gaussian error linear units ({GELU}s)},
  author  = {Hendrycks, Dan and Gimpel, Kevin},
  journal = {arXiv preprint arXiv:1606.08415},
  year    = {2016}
}

@article{ba2016layer,
  title   = {Layer normalization},
  author  = {Ba, Jimmy Lei and Kiros, Jamie Ryan and Hinton, Geoffrey E.},
  journal = {arXiv preprint arXiv:1607.06450},
  year    = {2016}
}

@inproceedings{fredrikson2015model,
  title     = {Model inversion attacks that exploit confidence information and basic countermeasures},
  author    = {Fredrikson, Matt and Jha, Somesh and Ristenpart, Thomas},
  booktitle = {Proc. ACM SIGSAC Conf. Comput. Commun. Secur. (CCS)},
  pages     = {1322--1333},
  year      = {2015}
}

@article{hwang_scores_2026,
  title   = {Scores know {Bob}'s voice: Speaker impersonation attack},
  author  = {Hwang, Chanwoo and Kim, Sunpill and Tan, Yong Kiam and Liu, Tianchi and Paik, Seunghun and Kim, Dongsoo and Soumik, Mondal and Aung, Khin Mi Mi and Seo, Jae Hong},
  journal = {arXiv preprint arXiv:2603.02781},
  year    = {2026}
}

@inproceedings{heigold2016endtoend,
  title     = {End-to-end text-dependent speaker verification},
  author    = {Heigold, Georg and Lopez-Moreno, Ignacio and Bengio, Samy and Shazeer, Noam},
  booktitle = {Proc. IEEE Int. Conf. Acoust. Speech Signal Process. (ICASSP)},
  pages     = {5115--5119},
  year      = {2016}
}

@article{li_model_2024,
  title   = {Model inversion attacks through target-specific conditional diffusion models},
  author  = {Li, Ouxiang and Hao, Yanbin and Wang, Zhicai and Zhu, Bin and Wang, Shuo and Zhang, Zaixi and Feng, Fuli},
  journal = {arXiv preprint arXiv:2407.11424},
  year    = {2024}
}

@article{lu_fgmia_2025,
  title   = {{FGMIA}: Feature-guided model inversion attacks against face recognition models},
  author  = {Lu, Ye and Wang, Shen and Zhu, Guopu and Zhang, Zhaoyang and Huang, Jiwu},
  journal = {IEEE Trans. Inf. Forensics Secur.},
  volume  = {20},
  pages   = {8465--8480},
  year    = {2025}
}

@article{kimiteam2025kimiaudio,
  title   = {{Kimi-Audio} technical report},
  author  = {{Kimi Team} and Ding, Ding and Ju, Zeqian and Leng, Yichong and Liu, Songxiang and Liu, Tong and Shang, Zeyu and Shen, Kai and Song, Wei and Tan, Xu and others},
  journal = {arXiv preprint arXiv:2504.18425},
  year    = {2025}
}

@article{kenny2007jfa,
  title   = {Joint factor analysis versus eigenchannels in speaker recognition},
  author  = {Kenny, Patrick and Boulianne, Gilles and Ouellet, Pierre and Dumouchel, Pierre},
  journal = {IEEE Trans. Audio Speech Lang. Process.},
  volume  = {15},
  number  = {4},
  pages   = {1435--1447},
  year    = {2007}
}

@inproceedings{mitsui2024pslm,
  title     = {{PSLM}: Parallel generation of text and speech with {LLM}s for low-latency spoken dialogue systems},
  author    = {Mitsui, Kentaro and Mitsuda, Koh and Wakatsuki, Toshiaki and Hono, Yukiya and Sawada, Kei},
  booktitle = {Findings Assoc. Comput. Linguist. (EMNLP)},
  pages     = {2692--2700},
  year      = {2024}
}

@inproceedings{morris2023text,
  title     = {Text embeddings reveal (almost) as much as text},
  author    = {Morris, John X. and Kuleshov, Volodymyr and Shmatikov, Vitaly and Rush, Alexander M.},
  booktitle = {Proc. Conf. Empir. Methods Nat. Lang. Process. (EMNLP)},
  pages     = {12448--12460},
  year      = {2023}
}

@inproceedings{nachmani2023spectron,
  title     = {Spoken question answering and speech continuation using spectrogram-powered {LLM}},
  author    = {Nachmani, Eliya and Levkovitch, Alon and Hirsch, Roy and Salazar, Julian and Asawaroengchai, Chulayuth and Mariooryad, Soroosh and Rivlin, Ehud and Skerry-Ryan, R. J. and Tadmor Ramanovich, Michelle},
  booktitle = {Proc. Int. Conf. Learn. Represent. (ICLR)},
  year      = {2024}
}

@inproceedings{nagrani2017voxceleb,
  title     = {{VoxCeleb}: A large-scale speaker identification dataset},
  author    = {Nagrani, Arsha and Chung, Joon Son and Zisserman, Andrew},
  booktitle = {Proc. Annu. Conf. Int. Speech Commun. Assoc. (Interspeech)},
  pages     = {2616--2620},
  year      = {2017}
}

@article{reynolds2000speaker,
  title   = {Speaker verification using adapted Gaussian mixture models},
  author  = {Reynolds, Douglas A. and Quatieri, Thomas F. and Dunn, Robert B.},
  journal = {Digit. Signal Process.},
  volume  = {10},
  number  = {1--3},
  pages   = {19--41},
  year    = {2000}
}

@inproceedings{snyder2018xvectors,
  title     = {X-vectors: Robust {DNN} embeddings for speaker recognition},
  author    = {Snyder, David and Garcia-Romero, Daniel and Sell, Gregory and Povey, Daniel and Khudanpur, Sanjeev},
  booktitle = {Proc. IEEE Int. Conf. Acoust. Speech Signal Process. (ICASSP)},
  pages     = {5329--5333},
  year      = {2018}
}

@inproceedings{pizzi_introducing_2022,
  title     = {Introducing model inversion attacks on automatic speaker recognition},
  author    = {Pizzi, Karla and Boenisch, Franziska and Sahin, Ugur and B{\"o}ttinger, Konstantin},
  booktitle = {Proc. Symp. Secur. Priv. Speech Commun. (SPSC)},
  pages     = {11--16},
  year      = {2022}
}

@inproceedings{tsaprazlis_voxguard_2026,
  title   = {{VoxGuard}: Evaluating user and attribute privacy in speech via membership inference attacks},
  author  = {Tsaprazlis, Efthymios and Lertpetchpun, Thanathai and Feng, Tiantian and Karimireddy, Sai Praneeth and Narayanan, Shrikanth},
  booktitle = {Proc. IEEE Int. Conf. Acoust. Speech Signal Process. (ICASSP)},
  pages   = {19042--19046},
  year    = {2026}
}

@inproceedings{variani2014dvector,
  title     = {Deep neural networks for small footprint text-dependent speaker verification},
  author    = {Variani, Ehsan and Lei, Xin and McDermott, Erik and Lopez Moreno, Ignacio and Gonzalez-Dominguez, Javier},
  booktitle = {Proc. IEEE Int. Conf. Acoust. Speech Signal Process. (ICASSP)},
  pages     = {4052--4056},
  year      = {2014}
}

@inproceedings{wan2018ge2e,
  title     = {Generalized end-to-end loss for speaker verification},
  author    = {Wan, Li and Wang, Quan and Papir, Alan and Lopez Moreno, Ignacio},
  booktitle = {Proc. IEEE Int. Conf. Acoust. Speech Signal Process. (ICASSP)},
  pages     = {4879--4883},
  year      = {2018}
}

@inproceedings{wang2023campp,
  title     = {{CAM++}: A fast and efficient network for speaker verification using context-aware masking},
  author    = {Wang, Hui and Zheng, Siqi and Chen, Yafeng and Cheng, Luyao and Chen, Qian},
  booktitle = {Proc. Annu. Conf. Int. Speech Commun. Assoc. (Interspeech)},
  pages     = {5301--5305},
  year      = {2023}
}

@inproceedings{wang2026hearsay,
  title   = {{HearSay} benchmark: Do audio {LLM}s leak what they hear?},
  author  = {Wang, Jin and Luo, Kaiwen and Lin, Liang and Wang, Weiliu and Chen, Yitian and Aloqaily, Moayad and Tang, Xuehai and Zhou, Zhenhong and Wang, Kun and Sun, Li and Wen, Qingsong},
  booktitle = {Findings Assoc. Comput. Linguist. (ACL)},
  pages   = {19312--19331},
  year    = {2026}
}

@article{xu2025qwen3omni,
  title   = {{Qwen3-Omni} technical report},
  author  = {Xu, Jin and Guo, Zhifang and Hu, Hangrui and Chu, Yunfei and Wang, Xiong and He, Jinzheng and Wang, Yuxuan and Shi, Xian and He, Ting and Zhu, Xinfa and others},
  journal = {arXiv preprint arXiv:2509.17765},
  year    = {2025}
}

@inproceedings{zhang2020secretrevealer,
  title     = {The secret revealer: Generative model-inversion attacks against deep neural networks},
  author    = {Zhang, Yuheng and Jia, Ruoxi and Pei, Hengzhi and Wang, Wenxiao and Li, Bo and Song, Dawn},
  booktitle = {Proc. IEEE/CVF Conf. Comput. Vis. Pattern Recognit. (CVPR)},
  pages     = {253--261},
  year      = {2020}
}

@inproceedings{zhang2023speechgpt,
  title     = {{SpeechGPT}: Empowering large language models with intrinsic cross-modal conversational abilities},
  author    = {Zhang, Dong and Li, Shimin and Zhang, Xin and Zhan, Jun and Wang, Pengyu and Zhou, Yaqian and Qiu, Xipeng},
  booktitle = {Findings Assoc. Comput. Linguist. (EMNLP)},
  pages     = {15757--15773},
  year      = {2023}
}

@article{zhu2026omnivoice,
  title   = {{OmniVoice}: Towards omnilingual zero-shot text-to-speech with diffusion language models},
  author  = {Zhu, Han and Ye, Lingxuan and Kang, Wei and Yao, Zengwei and Guo, Liyong and Kuang, Fangjun and Han, Zhifeng and Zhuang, Weiji and Lin, Long and Povey, Daniel},
  journal = {arXiv preprint arXiv:2604.00688},
  year    = {2026}
}

\end{document}